\documentclass[12pt]{amsart}
\usepackage{amsfonts}
\usepackage{amssymb}
\usepackage{sgame}
\usepackage{accents}
\usepackage{hyperref}
\usepackage{geometry}                
\usepackage[parfill]{parskip}    
\usepackage{graphicx}
\usepackage{ectasects}
\usepackage{palatino}
\usepackage{indentfirst}
\usepackage{tikz}
\usepackage{natbib}
\usepackage{epstopdf}
\usepackage{graphicx}
\usepackage{bbm}
\usepackage{amscd}
\usepackage{amsmath}
\usepackage{amsthm}
\usepackage{amsfonts}
\usepackage{amssymb}
\usepackage{accents}
\usepackage{indentfirst}
\usepackage{endnotes}
\theoremstyle{definition}

\theoremstyle{remark}

\theoremstyle{plain}

\theoremstyle{plain}
\newtheorem{lemma}{Lemma}
\theoremstyle{plain}
\newtheorem{result}{Result}
\theoremstyle{plain}
\newtheorem{definition}{Definition}
\theoremstyle{definition}
\newtheorem{proposition}{Proposition}
\theoremstyle{remark}

\allowdisplaybreaks

\renewcommand{\baselinestretch}{1.2}

\begin{document}
\title{The Asymptotic Cost of Complexity}

\author{Martin W. Cripps}
\thanks{Department of Economics, University College London; m.cripps@ucl.ac.uk. 
My thanks are due to  Tim Christensen and Deniz Kattwinkel for their advice on this material and to a seminar audience at Lancaster University for their comments.}

\date{\today }

\begin{abstract}
We propose a  measure of learning efficiency for non-finite state spaces. 
We characterize the ``complexity'' of a learning problem by the \emph{metric entropy} of its state space.
We then describe how learning efficiency  is determined by this measure of complexity.
This is, then, applied to two models where agents  learn high-dimensional states. 
\end{abstract}
\maketitle

\section{Introduction}

A  Bayesian agent who observes a sequence of independent identical signals from a statistical experiment will distinguish one parameter, or state, from another at an exponential rate.%
\footnote{\cite{Chernoff52} first determined this rate and a simple treatment can be found  in  \cite{CoverThomas91}.}
This rate has been called the ``asymptotic value of information'' (\cite{MoscariniSmith2002}) or the ``learning efficiency index'' (\cite{FrickII22}).
In these characterizations of the rate of learning,
 the complexity of what is being observed appears to play no role.
Whether there is a finite set or a continuum of signals in each state, the rate of learning is only determined by a Renyi divergence between the two signal distributions.
Here we show that in more general settings the rate of learning is  determined by the complexity of the sets of signals and states.
The appropriate measure of  complexity in this context is the \emph{metric entropy} of the state space.%
\footnote{This notion was first defined by \cite{kolmogorov59}.}
It is the metric entropy of the state space that determines the rate that an agent learns the  state.
This concept measures how rapidly the number of elements in a finite approximation to the state space must grow if the accuracy of the approximation is to improve.
Thus the metric entropy captures how easy it is for the agent to make good approximations to the underlying complexity of the environment.

In contrast to the finite state-space case, learning in larger spaces  is not exponentially fast.
We will show how this rate of learning  varies with the complexity of the state-space.
The slower rate of learning also changes the asymptotic demand for information.
\cite{MoscariniSmith2002} showed that the demand of information grew approximately 
logarithmically as its price converges to zero.
We will show that in larger state spaces, with slower rates of learning, the demand for information grows much faster than this as its price shrinks.

In larger state spaces it is  sensible to adjust the  definition of learning efficiency.
Instead of the Bayesian measure used before, we propose a minimax measure of learning efficiency.
The benefit of this minimax definition of learning efficiency is that it does not require a prior to be specified,
because selecting a prior can be difficult to do in models with larger spaces of states.
We do show, however, that our minimax definition of learning efficiency is consistent with the measures used for finite states.

These ideas are, then,  applied to two specific learning problems in economics:
The first is the problem of a manager allocating tasks to worker. We show that if the worker's skills become more complex to describe, then the manager needs more data on the worker's past performance to learn the worker's skills and optimally assign tasks.
The second is the problem of an agent exploring an unknown space of alternatives who is trying to determine the most profitable subset of alternatives.
This is a more demanding problem as it requires the agent to have a good estimate of small sets of the state space. Hence we show that getting close to optimal behavior can be slow if the profitability of the alternatives is very complex.

 The rate of learning efficiency for large state spaces has been much studied in the statistics literature. 
Some of the results described below were first established by \cite{LeCam86} and \cite{Birge86}, 
but we adopt the approach of \cite{Yang99} who use elegant information-theoretic arguments to get upper and lower bounds on learning rates.
The literature on non-parametric Bayesian models also considers rates of learning in large state spaces, see for example \cite{GhosalVDV17} .

\cite{MoscariniSmith2002} initiated the study of learning efficiency by giving a careful analysis of the rates of learning in a finite-state model. 
Their motivation was to develop an asymptotic ordering of the informativeness of experiments and an accurate expression for the demand for information when sampling costs are small.
This rate is also used as a benchmark for rates of social and common learning, see for example \cite{Harel21}.
Here we consider learning efficiency in  infinite-dimensional state spaces.
In this context a new criterion emerges that determines the rate of learning---metric entropy. 
We also provide a much cruder measure of the demand for information but do show demand grows much more rapidly as the price falls.

High-dimensional learning problems have tended to feature active/experimental models of learning.
A notable early contribution to this literature is \cite{Aghionetal91} who showed how an analytic function could be learnt relatively cheaply. 
In contrast, the details of more complex functions, such as Brownian paths, are much harder to learn from experimentation (for example, \cite{Callander11}).
Here we focus on a simpler sampling-based model of learning, but in this context we can quantify the rate that analytic functions and  Brownian paths are learnt---state spaces that include Brownian paths have a greater metric entropy than the spaces that include only analytic functions.

This paper is organized as follows.
Section 2 describes the two  examples of high-dimensional learning problems we study.
Section 3 gives the model, its notation, and our new definition of learning efficiency.
It then briefly visits the finite-state case to show how the definition is consistent with earlier work.
Section 4 starts by providing the definition of complexity that we will use and then two  results  of \cite{Yang99}
are presented that describe how complexity is related to learning efficiency.
Then our main result on learning efficiency and metric entropy is given.
In Section 5 we revisit the examples to  show how metric entropy plays a role in their solution.
Section 6 considers some further issues such as the demand for information.

\section{Examples of Learning Problems}
\label{lab}

In this section we give two examples of settings that fit the general model of learning that is described in Section \ref{mn}.

\subsection{Allocating Effort to Tasks}

 In the literatures on trade, production and labor it is common to consider models where there are a continuum of tasks/inputs/products and an agent must decide how to combine them.%
\footnote{See for example: \cite{Macdonald82}, \cite{Eaton02}, \cite{Freund22} .}
The following paragraphs outline such a model.
It is usually assumed that the agents know how each of the tasks/inputs/goods affect the eventual output that is produced. 
In contrast, we will consider the optimal decisions of an agent when this relationship is unknown.

There is a manager (she) who wants to maximize  output.
Output is produced by a newly-hired worker (he) who performs many tasks $x\in[0,1]$ and applies effort, or attention, to them.
A worker of type $\theta$ has productivity $\phi_\theta(x)$ for task $x$, where $\phi_\theta:[0,1]\rightarrow\mathbbm{R}_+$.
The manager regulates how much effort the worker allocates to each of the tasks:
she specifies an effort allocation function  $e:[0,1]\rightarrow\mathbbm{R}_+$.
The output from the task $x$ is the product of the worker's effort and his productivity,
 $q(x)=e(x)\phi_\theta(x)$.
The manager's problem, therefore, is to select an effort allocation $e(.)$ to maximize the total output.
Total output is a CES  aggregation of the outputs, $q(.)$, produced from the individual tasks. 
 Of course, the manager's choice of the worker's effort is constrained, $\int_{[0,1]}e\,d\nu\leq1$,  because the worker has a bounded amount of effort to distribute among the tasks.%
 \footnote{We use $\nu$ to denote the Lebesgue measure on $[0,1]$.}
 Therefore, the manger  chooses $e:[0,1]\rightarrow\mathbbm{R}_+$ to solve the problem:
$$
\max_{e(.)}\left[\int_{[0,1]} (e\phi_\theta)^{\frac{\sigma-1}{\sigma}}d\nu\right]^{\frac{\sigma}{\sigma-1}},
\qquad
\makebox{s.t.}
\qquad
\int_{[0,1]}e\,d\nu=1.
$$

Clearly, to make an optimal production decision 
it is necessary for the manager to  know the  function $\phi_\theta$.
Below, we will consider a manager who does not know the relationship between tasks and the worker's productivity.
We will show that it is the complexity (metric entropy) of the space of $\phi_\theta$\/'s that determines whether she makes a good or bad effort allocation decision for a given amount of data on $\phi_\theta$.
If $\phi_\theta$ were constant, or slowly varying, then lack of knowledge about the detailed shape of $\phi_\theta$ would not harm her much and so not accurately knowing $\phi_\theta$ will not  be
too  costly for the manager.
However, if $\phi_\theta$ is very variable at a fine scale and  the worker's abilities are very complex to describe, then it can take the manager a long time to take good decisions.

There are a number of different models of learning that one could apply in this setting.
The results we present are for a  manager who is able to learn about $\phi_\theta$ from the past history of tasks that the worker has performed.
To be precise, the manager observes a sequence $(x_1,\dots,x_t)$ of tasks the worker has undertaken in the past.
The worker is more likely to have done a task in the past if they are good/productive in that role, so 
this sequence is an iid sample from the density $f_\theta$.%
\footnote{ An alternative would be to model the manager's learning as an outcome of active experimenting in the allocation of tasks to the worker.
This is beyond the scope of the results in this paper.}

\subsection{Exploring a Space of Alternatives}

The second example is a model where an agent makes observations on a function in an attempt to find the best  alternative, or a set of good alternatives.
This is usually modeled as a problem of exploration/experimentation (see \cite{Callander11} or \cite{Liu24} for example).
The question of what kinds of functions are difficult to learn can also be
be posed in simpler learning settings, however.

There is an interval of alternatives $x\in[0,1]$. 
At each of these alternatives there is a likelihood $p_\theta(x)$ of achieving a particular outcome.
For example, an alternative, $x$, might be a technology choice and $p_\theta(x)$ is the probability of a successful innovation using that technology.
Or the alternative $x$ could be a subject and $p_\theta(x)$ the probability that subject $x$ contracts a disease and, therefore, would benefit from a treatment.
In this context it is without loss to suppose that $p_\theta(x)$ is a density on $[0,1]$.

There is an agent who has a limited budget to explore/treat these alternatives.
They are constrained to explore the subset of alternatives $D\subset[0,1]$, where $\nu(D)=d$ and $d<1$.%
\footnote{Recall that $\nu$ denotes Lebesgue measure on $\mathbbm{R}$.}
The agent's objective is to maximize the number of successful outcomes they achieve given their budget.
(The researcher wishes to maximize the expected number of successful innovations, or
the medic wishes to offer treatment to the subjects who are most likely to contract a disease.)
If the function $p_\theta$ were known, then the agent's problem is
$$
\max_{D\subset [0,1]} d^{-1}\int_{D}p_\theta \, d\nu,
\qquad
\nu(D)=d.
$$
(The term $d^{-1}$  is a normalization to ensure that even as $d\rightarrow0$ this optimization is well defined.)
The optimal choice $D$ in the problem above will be a set of intervals, if $p_\theta$ is continuous.
These intervals can be arbitrarily complex.

We study models where $p_\theta$ is unknown. 
How the agent learns about $p_\theta$ could again be modeled in many ways. 
The learning here is passive and the agent  observes an independent sample 
$(x_1,\dots,x_t)$ from the density $p_\theta$.
A rationalization for this is that in previous periods the technologies have generated innovations proportional to their likelihood.
Or, that in the past the subjects have 
contracted the disease with a probability that is proportional to their susceptibility.
We will show that the complexity of the set of functions $p_\theta$ determines how quickly a particular
$p_\theta$ is learnt.%
\footnote{A model of exploration pioneered by \cite{Callander11} instead assumes that in each period the agent selects a location $x$ and observes precisely $p_\theta(x)$.}

\section{The Model}
\label{mn}

In this section we provide the notation of the model, give a new definition of learning efficiency, and show that this consistent with the definition used in finite state spaces.

There is a decision maker (DM) who observes the outcomes of a sequence of identical statistical experiments.
The  experiment is described by a collection of densities $\mathcal{P}_\Theta:=\{p_\theta :\theta\in\Theta\}$, the index set $\Theta$ is compact.
These are densities with respect to the Lebesgue measure  $\nu$ on a compact interval $\mathcal{X}$.
The densities take values in a compact strictly positive interval of  $\mathbbm{R}$:   $p_\theta(x)\in[M^{-1},M]\subset \mathbbm{R}_+$ for all $(x,\theta)\in\mathcal{X}\times\Theta$.
This ensures the likelihood ratios are bounded $\frac{p_\theta(x)}{p_{\theta'}(x)}\leq M^2$, for all $x$, $\theta$ and $\theta'$.
The DM observes an iid sample $X_1,\dots,X_t\in\mathcal{X}^t$ from a ``true'' density $p_{\theta}$, for some $\theta\in\Theta$.
We will use $\mathbbm{P}_\theta$  to denote the probability measure on $\mathcal{X}^t$ induced by iid samples from the density $p_\theta$.
Expectations taken with respect to this measure are denoted $E_\theta$.

The parameter set $\Theta$ that indexes these densities describes the ``complexity'' of the state space.
The index set $\Theta$ could be finite or a subset of $\mathbbm{R}^m$, these are usually termed  ``parametric models''.
The index set $\Theta$ can also be a set of probability measures in which case we can think of $\mathcal{P}_\Theta$ and $\Theta$ as being isomorphic; these are termed ``non-parametric'' models.%
\footnote{Here $\Theta$ will be a subset of the set of all probability measures that are absolutely continuous relative to Lebesgue measure on $\mathbbm{R}$. $\Theta$ is complete and separable relative to the total variation metric, see \cite{GhosalVDV17} p.512 for example.}

We  now propose a  definition of  learning efficiency that is based on the rate that the minimax risk converges to zero. 
This measure is common in the statistics literature (for example \cite{Birge86}) and is consistent with the  definitions of learning efficiency in  \cite{MoscariniSmith2002} or \cite{FrickII22}.
Moreover, it extends these measures to  spaces $\Theta$ where it may be technically difficult to find a satisfactory prior.
Thus it is not a Bayesian measure.%
\footnote{It is possible to adopt Bayesian approach to modeling complexity, see for example \cite{GhosalVDV17} Chapter 8.
Again the metric entropy plays a role in determining the rate of convergence, but the prior also plays a role.}

The DM  observes the sample $X_1,\dots,X_t$ and must predict the value of $\theta$.
The DM's prediction is a measurable function $\hat\theta_t:\mathcal{X}^t\rightarrow\hat\Theta$.%
\footnote{Usually we will choose $\hat\Theta=\Theta$ but in some non-parametric models it may be necessary to have $\hat\Theta\subset\Theta$.}
The DM's costs from an incorrect decision are described by a cost function $c:\Theta^2\rightarrow \mathbbm{R}_+$,
where $c(\theta,\theta)=0$. 
We will choose $c(\theta,\hat\theta_t)$ so that
$$
c(\theta,\hat\theta_t)
=
h^2(p_\theta,p_{\hat\theta_t})
:=
\int(\sqrt{p_\theta}-\sqrt{ p_{\hat\theta_t}})^2\,d\nu.
$$
$h(p_\theta,p_{\hat\theta_t})$ is the Hellinger distance between the signal density, $p_\theta$, and the density of the prediction $p_{\hat\theta_t}$.%
\footnote{In the examples below we have other measures for the costs of errors. 
We will show how one can use rates for the Hellinger distance to get bounds on rates for other measures of costs.}
The rate at which $h^2(p_\theta,p_{\hat\theta_t})$ converges to zero conditional on a given $\theta$,
 captures the rate at which one value of $\theta$ is learned.
As $\theta$ is itself unknown,  the minimax cost measures a rate of learning independent of a particular $\theta$
\begin{equation}
C_t:=\inf_{\hat\theta_t\in S_t} \sup_{\theta\in\Theta} E_\theta\left( \ h^2(p_\theta,p_{\hat\theta_t}) \ \right).
\label{mm}
\end{equation}
($S_t$ is the set of all $\hat\theta_t$.)
We will use the rate that $C_t$ tends to zero to be  
our parameter-independent measure of the speed of learning. 

\begin{definition}
Let $r_t\rightarrow0$ as $t\rightarrow\infty$, we say $\mathcal{P}_\Theta$ has learning efficiency $r_t$ if 
$$
\limsup_{t\rightarrow\infty}\frac{C_t}{r_t}<\infty.
$$
\end{definition}

There are many sequences $(r_t)$ that will satisfy this condition.
It is difficult to  determine precisely how fast $C_t$ goes to zero and so bounds on these rates are usually studied.

We end this section by showing, in Proposition \ref{msfii}, that the definition of learning efficiency above is consistent with those of
\cite{MoscariniSmith2002} and \cite{FrickII22} for finite parameter spaces.

If there is a finite set of states  $\Theta=\{\theta_1,\theta_2,\dots,\theta_n\}$, then the rate at which parameter $\theta_i$ is distinguished from parameter $\theta_j$ is determined by  
$\lambda_{ij}$, where 
$$
\lambda_{ij}:=\log\min_{\kappa\in[0,1]}\int_{\mathcal{X}} p_{\theta_i}(x)^\kappa p_{\theta_j}(x)^{1-\kappa}\nu(dx)\leq0.
$$
As $\lambda_{ij}$ becomes closer to zero, it becomes harder to distinguish $\theta_i$ from $\theta_j$.
As learning is exponentially fast in this setting the learning efficiency of state $\theta_i$ is determined by the largest value of $\lambda_{ij}$
and the learning efficiency of the experiment as a whole is determined by $\lambda^*:=\max_{ij}\lambda_{ij}$.
The following proposition shows that this is also true for the minimax measure defined in (\ref{mm}).
Although this proposition is well-known, an outline of the   proof is supplied in the Appendix.

\begin{proposition}
If $\Theta=\{\theta_1,\dots,\theta_n\}$ and $\lambda^*=\max_{ij}\lambda_{ij}$, then $\mathcal{P}_\Theta$ has learning efficiency $e^{\lambda t}$ for all $\lambda\in(\lambda^*,0)$ and 
$e^{\lambda^*t+o(t)}
\leq
C_t
\leq 
n^2 e^{\lambda^*t+o(t)}$,
where $o(t)$ is a function that satisfies $\lim_{t\rightarrow\infty} o(t)/t=0$.
\label{msfii}
\end{proposition}

 \section{Learning Efficiency and Complexity}
 
 The first part of this section describes our measure of complexity and explains the role it plays in learning problems with non-finite state spaces.
 Then we describe the results of \cite{Yang99} who show how this measure of complexity determines learning efficiency.
 First there is a result that provides an upper bound on learning efficiency in terms of metric entropy.
 Then, then there is a lower bound on learning efficiency that will allow us to argue that in certain cases we have a good approximation to learning efficiency.
 In the final section we show how these two bounds can be combined to get a close relationship between metric entropy and learning efficiency.

\subsection{Complexity of the State Space}
\label{rff}
 
 As $\Theta$ grows from a finite to an infinite set, there will be two problems in trying to apply Proposition \ref{msfii}.
 First,  different states may have signal densities that are arbitrarily close  to each other
 and so  $\lambda^*$ will approach zero.
As a result, learning will not  be exponentially fast.
A second issue is that as the number of states, $n$, increases the upper bound on $C_t$ in Proposition \ref{msfii} becomes increasingly weak.
This is because the true hypothesis must be tested against infinitely many alternatives.
The solution to both of these problems is to cover the space of densities $\mathcal{P}_\Theta$  with a finite collection 
of $\varepsilon$-balls in a given distance/metric.

The first issue is solved by the introduction of these balls, because then it will be sufficient to distinguish a ball around the true $\theta$ from all other alternatives.
Thus, the rate at which the true state, $\theta$, is learned depends upon the rate at which the probability of states outside this ball converge to zero.
Of course, as this ball shrinks the learning will slow.

The second issue, of there being an infinite number of alternatives, is also helped by the $\varepsilon$-ball approach.
If the state space is totally bounded then there will be a finite number of balls that cover it.
Thus for a given $\varepsilon$ there will be only a finite number of alternatives to rule out in the upper bound.
However, for a precise estimate of the state it is necessary to let $\varepsilon\rightarrow0$.
When this occurs the number of balls covering the parameter space might  grow large very quickly
making the upper bound useless.
How the number of balls grows as $\varepsilon\rightarrow0$ is captured by the metric entropy and this will be our measure of the complexity of the parameter space.
Thus metric entropy/complexity captures how difficult it is to get accurate information on the fine detail of a state.

The first step in defining metric entropy is to define a measure of distance to be used.
This will be based on the  Kullback-Leibler (KL) divergence 
$$
K(p_\theta\Vert p_{\theta'}):=E_\theta\left(\log \frac{p_\theta}{p_{\theta'}}\right)=\int_{\mathcal{X}}p_\theta\log\frac{p_\theta}{p_{\theta'}}\,d\nu.
$$
We define the distance $d_K(\theta,\theta'):=\sqrt{K(p_\theta\Vert p_{\theta'}) }$; this is not a metric.
If the likelihood ratios of the densities are bounded, however, this is an  equivalent notion of distance to the Hellinger distance.%
\footnote{The KL divergence and the Hellinger distance satisfy the inequalities
$h(p_\theta,p_{\theta'})\leq d_K(\theta,\theta')\leq \sqrt{2}h(p_\theta,p_{\theta'})\left\Vert\frac{p_\theta}{p_{\theta'}}\right\Vert_\infty^{1/2}$ (see \cite{GhosalVDV17} Appendix B).
\label{rf}}
We also define  $\varepsilon$-ball centered at $p_\theta$ in the distance $d_K$
$$
B^K_\varepsilon(p_\theta)=\{q\in\mathcal{P}_\Theta:d_K(p_\theta,q)\leq\varepsilon\}.
$$

There are at least two ways in which we can use the balls $B^K_\varepsilon(p_\theta)$ to approximate a space.
You can cover the space with a minimal number of balls or you can pack the space with points that are at least $\varepsilon$ apart in the distance $d_K$.
This is made more precise in the following definition.

\begin{definition}
The $\varepsilon$-covering number, $N_d(\varepsilon)$, is the minimal number of $\varepsilon$-balls in the distance $d$ needed to cover $\mathcal{P}_\Theta$.
The $\varepsilon$-packing number $D_d(\varepsilon)$ is the maximal number of points in $\mathcal{P}_\Theta$ that are at least $\varepsilon$ apart in the $d$ distance. 
\end{definition}

The logarithm of the number of balls used in each of these approximations to a space determines two different notions of the space's entropy. 

\begin{definition}
The \emph{packing} entropy of $\mathcal{P}_\Theta$ is $\log D_d(\varepsilon)$,  
The \emph{covering} entropy of $\mathcal{P}_\Theta$ is $\log N_d(\varepsilon)$
\end{definition}

The covering number and the packing number are not identical but they are of the same order.
In particular, if the distance $d$ satisfies the triangle inequality,
 then it is easy to see that $
N_d(\frac{\varepsilon}{2})\geq D_d(\varepsilon)\geq N_d(\varepsilon)$.
Hence we will use the terminology \emph{Metric Entropy} to refer to both the packing and covering entropies.

Now we present a result that shows  the role metric entropy plays when a finite approximation of a state space is taken.
The result shows that  the distance between the true sample density $p_\theta(X^t)$ and its density under a finite approximation $q(X^t)$  is bounded above by the metric entropy and a factor that grows linearly in the sample size.
The distance between the two densities is measured using the Kullback-Leibler divergence 
$K( p_\theta(X^t)\Vert q(X^t))$.

To define this finite approximation, let $G_K(\varepsilon):=\{\theta_1,\dots,\theta_{N_K(\varepsilon)}\}$ 
be centers of the balls in the $\varepsilon$-cover $N_K(\varepsilon)$.
Fix  a prior that is uniform on the points $G_K(\varepsilon)$, then
under this prior the density of  the outcome $X^t=(X_1,\dots,X_t)$ is
$$
q(X^t):=\frac{1}{N_K(\varepsilon)} \sum_{\theta_j\in G_K(\varepsilon)}p_{\theta_j}(X^t).
$$
The following lemma gives an upper bound on how well $q$ approximates $p_\theta(X_t)$ in terms of its divergence.
This obviously  depends on
the number of points in the support of the finite prior and how close these points are to $\theta$.
The proof of this lemma is described in the Appendix.

\begin{lemma}[\cite{Yang99}]
$K( p_\theta(X^t)\Vert q(X^t)) \leq \log N_K(\varepsilon)+t\varepsilon$.
\label{sb}
\end{lemma}

\subsection{An Upper Bound on Learning Efficiency}

This section contains a simplification of result due to \cite{Yang99} that bounds the learning efficiency from above.
It shows how the upper bound in Proposition \ref{msfii} will change when the size of the state space grows.
As we have suggested in the discussion, the upper bound we find depends not on the number of states but on
the metric entropy of the state space.

The proof of this Proposition is given in the appendix.
The upper bound consists of two functions $\varepsilon^2$, which decreases as $\varepsilon\rightarrow0$, and $t^{-1}\log N_K(\varepsilon)$, that increases as $\varepsilon\rightarrow0$.
These reflect the two aspects of the problem outlined above.
First, as $\varepsilon$ decreases an $\varepsilon$-packing better approximates the state space, so any prediction based on this finer approximation will be better. 
Second, as $\varepsilon$ decreases the number of sets in an $\varepsilon$-cover increases so there are many more alternatives to exclude. This is a cost of complexity.
For a given value of $t$ this upper bound is minimized by trading off these two costs.

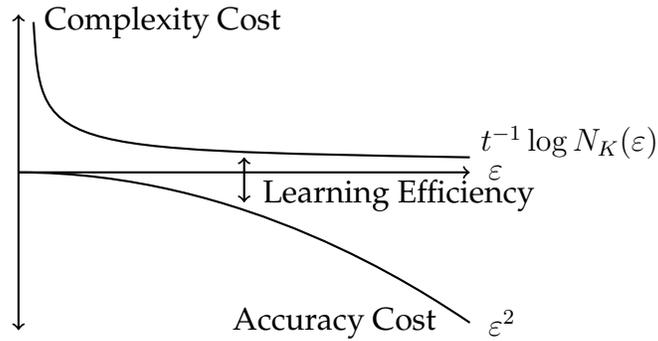
\begin{figure}[h]
\begin{center}
\begin{tikzpicture}[scale=1.0]
\draw [thick,->] (0,0) -- (6,0);
\node [right] at (6.1,0) {$\varepsilon$};
\draw [thick,->] (0,0) -- (0,2.1);
\draw [thick,->] (0,0) -- (0,-2.1);
\node [right] at (0.2,2) {Complexity Cost};
\node [right] at (2.7,-2) {Accuracy Cost};
\draw [thick] (0,0) parabola (6,-2);
\draw [thick] (0.2,2) .. controls (0.3,0.3) .. (6,0.2);
\node [right] at (6.1,-2) {$\varepsilon^2$};
\node [right] at (6,0.4) {$t^{-1}\log N_K(\varepsilon)$};
\draw [thick,->] (3,0) -- (3,0.2);
\draw [thick,->] (3,0) -- (3,-0.4);
\node [right] at (3.1,-0.3) {Learning Efficiency};
\end{tikzpicture}
\end{center}
\caption{The Upper Bound on Learning Efficiency}
\label{f2}
\end{figure}

The argument of the proof is based on the finite approximation that was introduced in the previous section.
It proposes a simple prediction $\tilde q_t$ of the true density that is based on this finite approximation and shows that the distance between the prediction and the true density 
$E_\theta(K(p_\theta(X^t)\Vert \tilde q_t(X^t))$ is bounded above by $t^{-1}K( p_\theta(X^t)\Vert q(X^t))$
using the convexity properties of the divergence.
Then the Lemma and the fact that $h^2\leq K$ gives the bound.

\begin{proposition}[Yang and Barron]
If $\mathcal{P}_\Theta$ is convex and $\varepsilon>0$ then 
\begin{equation}
\inf_{\hat\theta_t\in S_t}\sup_{\theta\in\Theta}
E_\theta\left( h^2(p_\theta , p_{\hat\theta_t})\right)
\leq \varepsilon^2+t^{-1}\log N_K(\varepsilon).
\label{yb1}
\end{equation}
\label{up}
\end{proposition}

The function $\log N_K(\varepsilon)$ is highly discontinuous, as $N_K(\varepsilon)$ takes integer values and becomes arbitrarily large as $\varepsilon\rightarrow0$. 
Thus finding the value of $\varepsilon$ that minimizes this bound for a given value of $t$ can be difficult.
This of lack continuity is suppressed in Figure \ref{f2}  which is only intended to illustrate the trade offs in the choice of $\varepsilon$.
In practice there are well-defined continuous approximations to $N_K(\varepsilon)$ which can be used to 
approximate the tightest upper bound on learning efficiency.
As $t$ increases the complexity costs shift downwards and the optimal choice of $\varepsilon$ shrinks---
this is how the complexity cost determines the rate of learning.

\subsection{A Lower Bound on Learning Efficiency}

It is far from obvious that the above upper bound is tight.
It might be the case that learning efficiency converges to zero much faster than this upper bound suggests.
Fortunately \cite{Yang99}, also give a lower bound that implies the upper does give a good approximation to learning efficiency in certain cases. This result is described below.

The proof of this lower bound, also presented in the appendix, is  based on information theoretic ideas.
The intuition for the lower bound follows from the following story.
Suppose that the DM is restricted to only pick its estimator of the true state from an $\xi$-packing set.
In this case, if the DM does not pick an estimator that is closest to the true state, then the DM experiences a
error cost of at least $\xi^2/4$.
Thus a lower bound on the DM's costs from this estimator is $\xi^2/4$ multiplied by the probability the DM makes this incorrect choice.
A lower bound on the probability of an incorrect choice is given by Fano's inequality
applied to the mutual information between the observed sample and the restricted set of states.
This gives the term in square brackets below.
Lemma \ref{sb} is then used to bound the mutual information term that appears in Fano's inequality.

\begin{proposition}[Yang and Barron]
If $\xi>0$ and $\varepsilon>0$, then 
\begin{equation}
\inf_{\hat\theta_t\in S_t}\sup_{\theta\in\Theta}
E_\theta\left( h^2(p_\theta , p_{\hat\theta_t})\right)
\geq 
\frac{\xi^2}{4}
\left[1-\frac{\log N_K(\varepsilon) + t\varepsilon^2 +\log 2}{\log N_K(\sqrt{2}M\xi)}\right]
.
\label{dd}
\end{equation}
\label{down}
\end{proposition}

At first sight it seems unclear how the lower bound relates to the upper bound of Proposition \ref{up}.
The following subsection shows how this can be done for a particular class of models.

\subsection{Learning Efficiency and Complexity}

In this section we show how the upper and lower bounds above can be combined to provide a good estimate of the learning efficiency.
This then shows that learning efficiency is determined by the metric entropy.
To do this it is necessary to be more precise about the class of models $\mathcal{P}_\Theta$. 
There are many results on the metric entropy of classes of functions, so in making this choice it is necessary to find a set for which a good approximations to the entropy exits.

We begin by describing a class of densities, $\mathcal{P}_\Theta^{\alpha,n,\lambda}$,
for which estimates of the metric entropy do exist.
The class $\mathcal{P}_\Theta^{\alpha,n,\lambda}$ consists of densities that are described by three parameters $(\alpha,n,\lambda)$.
They are functions that are $n-1$ times continuously differentiable and
they have an $n^{\rm th}$ derivative that satisfies a H\"older-continuity condition with parameters $\alpha\in(0,1]$ and $\lambda>0$.%
\footnote{This is sometimes called a generalized H\"older space and is closely related to the notion of a Sobolev space where the norm also controls the derivatives.}

\begin{definition}
Let $n\in\mathbbm{Z}_+$, $\alpha\in(0,1]$ and $\lambda\geq0$ be given.
Then $p_\theta\in \mathcal{P}_\theta^{\alpha,n,\lambda}$, if and only if: $p_\theta$ is a density on $[0,1]$, $p_\theta(x)\in[M^{-1},M]$, $p_\theta$ is $n-1$-times continuously differentiable with derivative bounded in absolute value by $\lambda$, and its $n^{\rm th}$ derivative, $p_\theta^{(n)}$, satisfies
$$
|p^{(n)}_\theta(x)-p^{(n)}_\theta(x')|\leq \lambda|x-x'|^\alpha,\qquad \forall x,x'\in[0,1].
$$
\end{definition}

As $n$ or $\alpha$ increase (and $\lambda$ decreases) the functions in this class will be more  smooth and vary less.
Thus the complexity a of function $p_\theta\in \mathcal{P}_\theta^{\alpha,n,\lambda}$ will be reduced and it becomes easier to learn.
The most complex functions in this class occur when $n=0$. 
For example, $\mathcal{P}_\theta^{\alpha,0,\lambda}$ contains the Lipschitz functions and densities that satisfy the H\"older condition $|p_\theta(x+h)-p_\theta(x)|\leq \lambda|h|^\alpha$.
The class of densities $P_\Theta^{\alpha,0,\lambda}$ grows to include more variable functions as $\alpha$ decreases from unity.
The class $\mathcal{P}_\theta^{\alpha,0,\lambda}$ with $\alpha<0.5$  is of particular interest, because it includes functions that are paths of Brownian motion, see \cite{revuz99} p.28.

The metric entropy of the class $\mathcal{P}_\theta(n,\alpha,\lambda)$ was first characterized by  \cite{kolmogorov59} who used the $L_\infty$ norm.
Then \cite{Clements63} showed that these properties also hold in the $L_1$  norm and \cite{Lorentz66} for the $L_p$ norms.
These all show that 
$$
\log N_p(\varepsilon)=O(\varepsilon^{-\frac{1}{n+\alpha}}),
\qquad
p\in[1,\infty].
$$
As $\varepsilon\rightarrow0$  the entropy grows  exponentially with a factor determined by $r=n+\alpha$. 
As $r=n+\alpha$ shrinks so the exponential factor grows hence so too does the complexity of the space being considered.

The following result characterizes the learning efficiency for the class $\mathcal{P}_\Theta^{\alpha,n,\lambda}$.
It shows that this is determined  by the parameter $r$ that controls the metric entropy of the space.
Thus it is the metric entropy that determines learning efficiency.
The key step in the proof is to show that the metric entropy of this class in the $K$-distance is well approximated by 
its metric entropy in the $L_2$ norm.
Once this is done it is necessary to optimize the approximate upper and lower bounds to achieve the bounds stated in the proposition.

\begin{proposition}
For the class of densities $\mathcal{P}_\Theta^{\alpha,n,\lambda}$ with $r=\alpha+n$ there exists constants $C',C$ dependent on $\alpha$, $M$ and $\lambda$ such that
\begin{equation}
C' t^{\frac{-2r}{1+2r}}-O(t^{-1})
\leq
\inf_{\hat\theta_t\in S_t}\sup_{\theta\in\Theta}
E_\theta\left( h^2(p_\theta , p_{\hat\theta_t})\right) 
\leq 
C t^{\frac{-2r}{1+2r}}
\end{equation}
\label{tig}
\end{proposition}

Thus as class of densities $\mathcal{P}_\Theta^{\alpha,n,\lambda}$ grows to include more variable/wiggle-ly functions ($r=\alpha+n$ shrinks), so it takes longer to achieve a given level of certainty about the state.
The best rate achievable is $t^{-1}$ for infinitely differentiable functions and the  rate becomes arbitrarily bad as $r$ gets small. It is almost $t^{-1/2}$ for $(\alpha,n)=(\frac{1}{2},0)$ when the state includes Brownian paths.
The Lipschitz parameter $\lambda$ does  play a role, but its effect is of a logarithmic order and it determines the size of the bounding constant.

\section{Complexity and Optimal Decisions}

In this section we return to the examples of Section \ref{lab} and show how complexity impacts these optimization problems.

\subsection{Allocating Effort to Tasks} 

There is a manager who wants to correctly
estimate a worker's ability in order to allocate tasks appropriately.
The output that gets produced increases as the manager's estimate of the worker's ability improves,
thus there is a direct connection between the quality of the manager's estimate and their profit.
It is not surprising, therefore, that the speed of learning or learning efficiency, determines how well the manager does at allocating tasks under imperfect information.
What is interesting is that the complexity of workers (as measured by the metric entropy of their state space) exactly determines the rate of convergence.
If workers have more variable abilities and this variation can occur at a very fine scale ($n=0$ $\alpha<1$), then it will take the manager much longer to get to a good allocation of tasks.
Whereas, if the worker's abilities are relatively simple to describe, then the manager can make good allocations of tasks with relatively little information.

When the worker's type, $\phi_\theta$, is known it is simple to see what the solution to the manager's effort-allocation problem should be.
Recall that  the manger  chooses $e:[0,1]\rightarrow\mathbbm{R}_+$ to solve:
$$
\max_{e(.)}\left[\int_{[0,1]} (e\phi_\theta)^{\frac{\sigma-1}{\sigma}}d\nu\right]^{\frac{\sigma}{\sigma-1}},
\qquad
\makebox{s.t.}
\qquad
\int_{[0,1]}e\,d\nu=1.
$$
Let us transform the worker's task-productivity relationship into a density $f_\theta(x):=\frac{\phi_\theta(x)^{\sigma-1}}{\int\phi_\theta^{\sigma-1}dx}$,
then the  maximization can be rewritten as
$$
\max_{e(.)}
C(\phi_\theta)\left[\int_{[0,1]} e^{1-\frac{1}{\sigma}}f_\theta^{\frac{1}{\sigma}}\,d\nu \right]^{\frac{\sigma}{\sigma-1}},
\qquad
\makebox{s.t.}
\qquad
\int_{[0,1]}e\,d\nu=1.
$$
where $C(\phi)^{\sigma-1}:=\int\phi^{\sigma-1}dx$.
With this transformation $e(.)$ and $f_\theta(.)$ are both densities on $[0,1]$ and output is  a decreasing function of the Renyi divergence between  them.%
\footnote{The Renyi divergence for two densities is $D_\sigma(p\Vert q):=\frac{1}{\sigma-1}\log\int p^\sigma q^{1-\sigma} \, d\nu$.
\label{re}}
This divergence is minimized by $e=f_\theta$ and so output is maximized by $e=f_\theta$.
If the manager knows the function $f_\theta$, then her 
optimal choice of effort is $e=f_\theta$.
Clearly, to make an optimal production decision 
it is necessary for the manager to  know the  function $f_\theta$ completely.
Here we will treat $f_\theta$ as being  unknown and assume that the manager does know the maximum feasible output the worker can produce $C(\phi_\theta)$.
Hence we focus on the high-dimensional aspects of the manager's problem, that is, how to learn how  to combine effort and tasks to achieve this maximum.%

The proposition below considers a manager who has to determine the effort allocation $e:[0,1]\rightarrow\mathbbm{R}_+$ of an unknown worker.
The worker's abilities are described by an unknown density $f_\theta$, although it is known that $f_\theta\in \mathcal{P}_\Theta^{\alpha,n,\lambda}$.
The manager does not know $f_\theta$ but does observe a sample $X^t=(X_1,\dots,X_t)\in[0,1]^t$ from the worker's true ability-density $f_\theta$.
The manager  uses this sample to make a prediction $e_t:[0,1]^t\rightarrow  \mathcal{P}_\Theta^{\alpha,n,\lambda}$ of the worker's actual ability.
The following result, with a proof in the appendix, shows that the manager's minimax expected payoff converges to its optimal value (of unity) at a rate that is determined by the metric entropy of the space $\mathcal{P}_\Theta^{\alpha,n,\lambda}$.
The result given in the proposition follows quite simply as we can use the Hellinger distance to bound the Renyi entropy that
is the manager's objective in the effort allocation optimization.
As  (\ref{tig}) gives bounds on the Hellinger distance combining these two bounds gives the result.

\begin{proposition}
For the class of densities $\mathcal{P}_\Theta^{\alpha,n,\lambda}$ with $r=\alpha+n$  and $\beta\geq\frac{1}{3}$, 
$$
C't^{\frac{-2r}{1+2r}}+O(t^{-1})
\leq
1-\sup_{e_t}\inf_{f_\theta\in \mathcal{P}_\Theta^{\alpha,n,\lambda}}E_\theta\left( \left(\int_{[0,1]} e_t^{1-\beta}f_\theta^\beta\,d\nu\right)^{\frac{1}{1-\beta}}\right)
\leq
\frac{C}{1-\beta}t^{\frac{-2r}{1+2r}}.
$$
\label{lo}
\end{proposition}

Thus the rate at which optimal behavior is attained is much slower when $n=0$ than when $n>0$.
Thus if the set $\mathcal{P}_\Theta^{\alpha,n,\lambda}$ includes paths of Brownian motion it can take
a very long time indeed to get close to optimal behavior.

\subsection{Exploring a Space of Alternatives}

In this section there is  an optimization that depends on the fine detail of the unknown function $f_\theta$ in an extreme 
way. In the task-allocation example above, it is important to learn the places where $f_\theta$ is large so that more effort is allocated to them.
However, as output is an integral of the individual tasks, it is possible to get close to optimal output 
even when the prediction is badly wrong on a small subset of tasks.
In the case of exploration it is only the alternatives that maximize $f_\theta$ that matter and this may be a small subset of all alternatives.
In contrast to the task-allocation problem, if the prediction of $f_\theta$ is badly wrong on a small set it is possible that the DM's payoff is badly approximated.
Thus although the metric entropy determines the rate at which learning proceeds and a similar rate applied to the optimal payoff in the  task-allocation problem.
Here we will find that the rate the behavior in the exploration problem approaches optimality can be much slower, (although it is still determined by the metric entropy). 

There is a set of alternatives $x\in [0,1]$ and a density $p_\theta(x)$ that determines the likelihood of a success at alternative $x$.
The Decision Maker (DM) can choose a measurable subset 
$D\subset [0,1]$ of alternatives to explore.
Let $\mathcal{D}$ be the collection of Lebesgue-measurable subsets of $[0,1]$ satisfying $\nu(D)=d$.
She has access to a sample $X^t=(X_1,\dots,X_t)\in I^t$ from the true density but is otherwise uninformed about the true function $p_\theta$.
A strategy for the DM is a function $D_t:[0,1]^t\rightarrow \mathcal{D}$
to maximise
$$
d^{-1}\int \mathbbm{1}_{D_t(X^t)}p_\theta\, d\nu,
$$
where $\mathbbm{1}_{D_t(X^t)}$ is the indicator function for the set $D_t(X^t)$.
Nature chooses $p_\theta\in\mathcal{P}_\Theta^{\alpha,n,\lambda}$ to make this problem as difficult as possible for the DM. Hence, the DM's objective is
$$
\sup_{D_t}\inf_{p_\theta\in\mathcal{P}_\Theta^{\alpha,n.\lambda}} E_\theta\left(\int \mathbbm{1}_{D_t(X^t)}p_\theta\, d\nu\right).
$$

We now prove a result that shows that again it is the metric entropy that determines the asymptotic behavior of this maximal payoff. 
A benchmark that we will measure this asymptotic behavior against is  the  achievable payoff for the DM if they actually knew the state $\theta$,
$$
M_\theta:=\max_{D\in\mathcal{D}}d^{-1} \int \mathbbm{1}_{D}p_\theta\, d\nu.
$$
This is clearly an upper bound on the DM's payoff thus the interest is in how quickly the DM's strategy gets close to achieving the payoff $M_\theta$.

\begin{proposition}
For the class of densities $\mathcal{P}_\Theta^{\alpha,n,\lambda}$ with $r=\alpha+n$ there exists constants $C',C$ dependent on $\alpha$, $M$, $\lambda$ and 
there is a strategy $\hat D_t$ for the decision maker so that for all 
$p_\theta\in\mathcal{P}_\Theta^{\alpha,n,\lambda}$
\begin{align*}
E_\theta\left( \, 
M_\theta -d^{-1}\int \mathbbm{1}_{\hat D_t}p_{\theta} \, d\nu
 \, \right)
 &\leq
 d^{-1}C t^{\frac{-\alpha^2}{(1+2\alpha)(1+\alpha)}},
 \qquad n=0;
 \\
 E_\theta\left( \, 
M_\theta -d^{-1}\int \mathbbm{1}_{\hat D_t}p_{\theta} \, d\nu
 \, \right)
 &\leq
 d^{-1}C t^{\frac{-r}{2(1+2r)}},
 \qquad 
 n>0.
\end{align*}
\label{fgg}
\end{proposition}

This result tells us  little about the solution to active exploration problems of \cite{Callander11} and \cite{Liu24} where the DM selects which points $x$ to sample and how much to explore.
One thing it does do is it provides a benchmark that we can compare to the equilibrium behavior in these more complex models.
One feasible exploration strategy for an agent is to sample randomly from $p_\theta$.
The above tells us how effective this will be at maximizing the DM's payoff and how important the metric entropy of the space is.

\section{Extensions and Further Issues}

Here we discuss some further aspects of our model of learning efficiency. 
We begin by considering what can be said about the costs and benefits of acquiring information.
Then, we consider the Bayesian approach to modeling learning efficiency. 
And finally we consider 
how these results perform when the state space $\Theta$ is finite dimensional.

In this setting the rate of learning is polynomial not exponential. This implies that the asymptotic  benefit of an extra piece of data is quite different from that found in  \cite{MoscariniSmith2002}.
In particular, as the price per unit of information shrinks, the demand of information grows much more rapidly.
As an example, and a crude approximation to the full problem, consider a DM who must decide how many units of data to commit to sample before deciding how to allocate tasks to an employee.
Assuming each independent data point in the sample can be brought at the price $\pi$, 
their information costs of committing to sample $t$ points would be $\pi t$. 
The benefit from this information (assuming a minimax objective) could be approximated using the
 bounds given in Proposition \ref{lo}. 
 Their optimization would therefore be  approximated by the  solution to the problem
 $\max_t 1-Kt^{-\frac{2r}{1+2r}}-\pi t$.
 This would yield  a demand for information (for $\pi$ sufficiently small) of 
 $$
 t=\left(\frac{2rK}{1+2r}\frac{1}{\pi}\right)^{\frac{1}{2}\frac{1+2r}{1+3r}}.
 $$
 Thus as $\pi$ becomes small the demand for information grows much faster than the logarithmic expression
of the finite-state case, \cite{MoscariniSmith2002}.

A Bayesian approach to this question would require the decision maker to put a prior on the space $\Theta$.
In the infinite-dimensional case ensuring this prior has a large enough support is non-trivial and was
 resolved by \cite{Schwartz65}.
 However, the rate at which the posteriors concentrate on the true parameter $\theta\in\Theta$ is still affected by the prior weight on a neighborhood of $\theta$ and how this weight behaves as the neighborhood shrinks. 
 Thus any measure of learning efficiency would be dependent on an assumption about the prior.
In spite of this issue, the metric entropy  is the main determinant of convergence rates  in Bayesian approaches to this problem.
This can be seen by the theory of posterior contraction rates in non-parametric models: \cite{GhosalVDV17} Chapter 8.

We have considered two extreme cases of the state sets: $\Theta$ is finite or $\Theta$ is infinite dimensional. There is obviously an important intermediate case where $\Theta$ is a totally bounded subset of $\mathbbm{R}^d$ and is finite dimensional.
This case includes most parametric models of learning.
Here the metric entropy is not a polynomial in $\varepsilon$ (the size of the balls), but logarithmic.
For example, a rectangle in $\mathbbm{R}^m$ has metric entropy of order $-m\log\varepsilon$ (see for example \cite{Wainwright19} Chapter 5).
The upper and lower bounds above still apply to this case.
If the upper bound of Proposition \ref{up} is calculated, it is of the correct order for this model: $t^{-1}$.
However, in calculating the lower bound of Proposition \ref{down} in this case one gets something that is far from the upper bound.
A more sophisticated argument is necessary to get a tight lower bound.
This uses local approximations to the state space rather than the global one.

\section{Conclusion}

We use the results of \cite{Yang99} to show how the metric entropy determines the asymptotic value of information and rates of learning.
This can be applied to many models where the unknown state is a function. We give two examples of this.


\renewcommand{\baselinestretch}{1}
\normalsize

\bibliographystyle{econometrica}

\bibliography{mwc}
\addcontentsline{toc}{section}{~~~~References}

\setcounter{section}{0}
\setcounter{equation}{0}
\renewcommand{\thesection}{\Alph{section}}	
\renewcommand{\theequation}{\thesection.\arabic{equation}}	

\section{Appendix}

\subsection{Proof of Proposition \ref{msfii}}
\begin{proof}
If $\Theta=\{\theta_1,\theta_2,\dots,\theta_n\}$, then
$$
C_t=\inf_{\hat\theta_t\in S_t} 
\max_i  \sum_{j\not=i}h^2(\theta_i,\theta_j)\mathbbm{P}_{\theta_i}(\hat\theta_t=\theta_j).
$$
Suppose that $h^2(\theta_i,\theta_j)\in[\underline{c}, \bar c]$ for all $i\not=j$ and that
$B_t:=\max_i  \mathbbm{P}_{\theta_i}(\hat\theta_t\not=\theta_i)$ is the maximum probability of $\hat\theta_t$ making an error in prediction.
Then $
 \underline{c}\inf_{\hat\theta_t\in S_t}B_t\leq C_t\leq \bar c \inf_{\hat\theta_t\in S_t}B_t$.
 If we focus only on the probability of error in two states we have
 $$
 \inf_{\hat\theta_t\in S_t}\max\{ \mathbbm{P}_{\theta_i}(\hat\theta_t=\theta_j),\mathbbm{P}_{\theta_j}(\hat\theta_t=\theta_i)\}
 \leq
 \inf_{\hat\theta_t\in S_t} B_t,
 \qquad \forall i\not=j.
 $$
The LHS is a prediction's maximum probability of error in an experiment with only two states.
Restricting to two states will make the infimum smaller and result in a different prediction process.

Let  $\hat\theta^{(ij)}_t$ be the prediction that minimizes the expected probability of error in an experiment with these two states.
The strategy that achieves this is described in Chapter 12 of  \cite{CoverThomas91}.
There it is also shown that rate that the expected probability of error converges to zero is proportional to the rate the maximum probability of error converges to zero.
That is:
\begin{align}
\inf_{\hat\theta_t\in S_t}\max\{ \mathbbm{P}_{\theta_i}(\hat\theta_t=\theta_j),\mathbbm{P}_{\theta_j}(\hat\theta_t=\theta_i)\}
 &= b\max\{ \mathbbm{P}_{\theta_i}(\hat\theta^{(ij)}_t=\theta_j),\mathbbm{P}_{\theta_j}(\hat\theta^{(ij)}_t=\theta_i)\}
\nonumber
 \\
 &=e^{\lambda_{ij} t+o(t)},
 \label{kk}
 \\
\lambda_{ij}&:=\log \min_{\kappa\in[0,1]}\int p_{\theta_i}(x)^\kappa p_{\theta_j}(x)^{1-\kappa}dx.
\nonumber
\end{align}

The parameters $\lambda_{ij}$ are the learning efficiency index for the two-state experiment (ignoring the terms of lower orders).
Combining the inequalities above we get 
$$
e^{\lambda_{ij} t+o(t)}=\inf_{\hat\theta_t\in S_t}\max\{ \mathbbm{P}_{\theta_i}(\hat\theta_t=\theta_j),\mathbbm{P}_{\theta_j}(\hat\theta_t=\theta_i)\}
\leq 
\inf_{\hat\theta_t\in S_t} B_t
\leq
C_t.
$$
As this holds for all $ij$ and  $\lambda^*=\max_{ij}\lambda_{ij}$, this then gives the claimed lower bound on $C_t$, that is, $e^{\lambda^*t+o(t)} \leq C_t$.

Now derive an  upper bound on $C_t$  that  depends on the number of states.
For each of the pairs $(i,j)$ there is an optimal prediction $\hat\theta^{(ij)}_t$ that achieves the
bound (\ref{kk}).
We use these $n(n-1)$ predictions and to define a new prediction strategy $\theta^*_t$.
This picks a state only if it passes all of its pairwise tests, that is,
$\theta^*_t=\theta_i$ if $\hat\theta^{(ij)}_t=\theta_i$ for all $j\not=i$. (We will enlarge the set of predictions to allow for an empty prediction to be made if no $\theta_i$ satisfies this condition.) 
By the usual exclusion bound
$$
\mathbbm{P}_{\theta_i}(\theta^*_t=\theta_i)
\geq 1-\sum_{j\not=i} \mathbbm{P}_{\theta_i}(\theta^{(ij)}_t=\theta_j).
$$
So by the equality (\ref{kk}),
$\mathbbm{P}_{\theta_i}(\theta^*_t\not=\theta_i)
\leq \sum_{j\not=i} e^{\lambda_{ij}t+o(t)}$.
Although $\theta^*$ is not an optimal strategy it certainly provides an upper bound on the costs from following an optimal strategy. 
A substitution into $B_t$ then gives
$$
C_t\leq \bar c \max_i\sum_{j\not=i} e^{\lambda_{ij}t+o(t)}\leq 
\bar c \sum_{ij \atop j\not=i} e^{\lambda_{ij}t+o(t)}
\leq 
n^2 e^{\lambda^*t+o(t)}
$$
Suppose $\lambda\in(\lambda^*,0)$ and 
combine this upper bound with the previous lower bound above we have shown.
Taking logs and dividing by $t$ then gives
$$
e^{t[\lambda^*-\lambda+t^{-1}o(t)]}
\leq
\frac{C_t}{e^{-\lambda t}}
\leq 
n^2e^{t(\lambda^*-\lambda+t^{-1}o(t)]},
$$
which implies $C_t/e^{-\lambda t}\rightarrow 0$ and $\mathcal{P}$ has learning efficiency $e^{-\lambda t}$.
 \end{proof}

\subsection{Proof of Lemma \ref{sb}}
\begin{proof}
For any $\theta_j\in G_K(\varepsilon)$,  we have 
\begin{align*}
K( p_\theta(X^t)\Vert q(X^t))
&=
E_\theta\left(\log\frac{p_\theta(X^t)}{N_K(\varepsilon)^{-1}\sum_{\theta_j\in G_K(\varepsilon)}p_{\theta_j}(X^t)}\right)
\\
&\leq
\log N_K(\varepsilon)+
E_\theta\left(\log\frac{p_\theta(X^t)}{p_{\theta_j}(X^t)}\right)
\\
&=\log N_K(\varepsilon)+tK(p_\theta\Vert p_{\theta_j})
\end{align*}
Choose $\theta_j\in G_K(\varepsilon)$ to attain the minimum of $K(p_\theta\Vert p_{\theta_j})$. This is less than $\varepsilon$ by the definition of $G_K(\varepsilon)$, so we get the inequality in the Lemma.
\end{proof}

\subsection{Proof of Proposition \ref{up}}

\begin{proof}
We begin by constructing a prediction $\tilde\theta_t$ of $\theta$.
There is a uniform prior on $G_K(\varepsilon)$.
As above, $q(X^t)$ denotes the density generated by this uniform prior.
Also, let $q(X_{t+1}|X^t)$ be the Bayesian prediction of the density of $X_{t+1}$ 
conditional on the history $X^t$ and the uniform prior on $G_K(\varepsilon)$.
Thus $q(X_{t+1}|X^t)$ is a Bayesian's best guess about the density of $X_{t+1}$ given this prior.
Now define a new density
$$
\tilde q_t(x) := \frac{1}{t}\sum_{m=0}^{t-1} q(x|X^m).
$$
This is the average of these predictive densities over the history $X^{t-1}$.
It is dependent on $X^t$ but this is suppressed in the notation
The convexity of $\mathcal{P}$ ensures that this prediction corresponds to a potential value of $\tilde\theta$.
(Otherwise it might be necessary to enlarge the set of admissible predictions.)
Now 
\begin{align*}
E_\theta \left(K(p_\theta\Vert \tilde q_t)\right)
&=
E_\theta K\left(p_\theta(.)\left\Vert t^{-1}\sum_{m=0}^{t-1} q(.|X^m) \right. \right)
\\
&\leq
E_\theta \frac{1}{t}\sum_{m=0}^{t-1} K(p_\theta(.)\Vert  q(.|X^m) )
\\
&=
\frac{1}{t}\sum_{m=0}^{t-1} E_\theta  K(p_\theta(X_{m+1})\Vert  q(X_{m+1}|X^m) )
\\
&=
\frac{1}{t}\sum_{m=0}^{t-1} E_\theta  \log\frac{p_\theta(X_{m+1})}{  q(X_{m+1}|X^m) }
\\
&=
\frac{1}{t} E_\theta  \log\frac{p_\theta(X^t)}{  q(X^t) }
=
\frac{1}{t}K(p_\theta(X^t)\Vert q(X^t))
\end{align*}
The expectation $E_\theta$ above applies to the conditioning variables $X^m$ used in producing the density $\tilde q_t$. The inequality follows from the convexity of the divergence.

The upper bound in Lemma \ref{sb} applies to the final expression (given the construction of  $q$), 
so $K(p_\theta(X^t)\Vert q(X^t))\leq \log N_K(\varepsilon)+ t\varepsilon^2$.
The upper bound here is independent of $\theta$ and holds for a particular prediction strategy 
$\tilde q_t$.
Thus taking the infimum will only reduce the expected costs and not violate this upper bound.
Thus we have 
$$
\inf_{\hat\theta_t\in S_t}\sup_{\theta\in\Theta}
E_\theta\left( K\left(p_\theta \Vert p_{\hat\theta_t}\right)\right)
\leq \varepsilon^2+t^{-1}\log N_K(\varepsilon).
$$
However the KL divergence satisfies the inequality
$h^2(p_\theta,p_{\theta'})\leq K(p_\theta\Vert p_{\theta'}) $, \cite[Appendix B]{GhosalVDV17}.
Hence we have the claimed result.
\end{proof}

\subsection{Proof of Proposition \ref{down}}

\begin{proof}
Let $\varepsilon>0$ and $\xi>0$ be given.
$\mathcal{D}_h(\xi)\subset\Theta$ is a largest $\xi$-packing  set (using the distance $h$) in $\Theta$
and  $D_h(\xi)$ is its cardinality. 
We will use $\mathbbm{P}_w$ to denote the measure on states of the world induced by a uniform prior on $\mathcal{D}_h(\xi)$. 
Under $\mathbbm{P}_w$ the histories $X^n$ have the density $p_w(X^t):=\sum_{\theta\in \mathcal{D}_h(\xi)} b p_\theta(X^t)$
, where  $b:=D_h(\xi)^{-1}$.
We will use $\theta_w$ to denote the random variable $\theta$ when it is generated from the measure  $\mathbbm{P}_w$.

Let $\hat\theta$ be an arbitrary point in $\Theta$ and let $\tilde\theta(\hat\theta)$ be a point in $\mathcal{D}_h(\xi)$ that is closest to $\hat\theta$ (if there are many pick one).
Then for any $\theta\in\mathcal{D}_h(\xi)$ with $\theta\not=\tilde\theta(\hat\theta)$
$$
\xi\leq h(\theta,\tilde\theta(\hat\theta))\leq h(\theta,\hat\theta)+h(\hat\theta,\tilde\theta(\hat\theta))\leq 2 h(\theta,\hat\theta).
$$
(As $h$ satisfies the triangle inequality.)
Thus we can deduce that for all $\hat\theta\in \Theta$  and all $\theta\in\mathcal{D}_h(\xi)$ with $\theta\not=\tilde\theta(\hat\theta)$
: $h(\theta,\hat\theta)\geq \xi/2$.
So, if a prediction of $\theta$ that is restricted to lie in $\mathcal{D}_h(\xi)$ does not pick this one closest point the cost (in terms of $h$) of this error is at least $\xi/2$.

The first step is to bound the min max expression from below by  
allowing nature to test estimators by only picking a $\theta\in\mathcal{D}_h(\xi)$.
There will only be a finite number of ways nature can test any prediction, so this allows a simple lower bound on the probability there is a big prediction error.
Given the sample $X^t$, $\hat\theta_t$ is the predictor of $\theta$ and  $\tilde\theta(\hat\theta_t)$ is a point in $\mathcal{D}_h(\xi)$ closest to this prediction.
Then, for any $\hat\theta_t$
\begin{align*}
\sup_{\theta\in\Theta} \mathbbm{P}_\theta\left( h(\theta,\hat\theta_t)\geq \xi/2\right)
&\geq
\max_{\theta\in \mathcal{D}_h(\xi)} \mathbbm{P}_\theta\left( h(\theta,\hat\theta_t)\geq \xi/2\right)
\\
&=
\max_{\theta\in \mathcal{D}_h(\xi)} \mathbbm{P}_\theta\left(\tilde\theta(\hat\theta_t)\not=\theta \right)
\\
&\geq
\sum_{\theta\in \mathcal{D}_h(\xi)} b\mathbbm{P}_\theta\left( \theta\not=\tilde\theta(\hat\theta_t) \right)
\\
&=
\mathbbm{P}_w\left( \theta_w\not=\tilde\theta(\hat\theta_t) \right).
\end{align*}
Above, the first inequality follows as the supremum is taken over a restricted set.
The second inequality replaces a maximum  with a convex combination.
The final equality recalls that the measure $\mathbbm{P}_w$ is generated from a uniform distribution over $\mathcal{D}_h(\xi)$.
The above expression has lower bounded a supremum over $\theta$ relative to the original measure, $\mathbbm{P}$, with  the  probability  that $\tilde\theta(\hat\theta_t)$ does not predict one of the finite set of states under the measure $\mathbbm{P}_w$.
This permits the next step of the argument.

The second step of the proof is to use Fano's inequality \cite[p. 38]{CoverThomas91} to provide a lower bound on $\mathbbm{P}_w\left( \theta_w\not=\tilde\theta(\hat\theta_t) \right)$. It provides a lower bound on the probability of 
making an incorrect prediction about a discrete random variable $\theta_w\in \mathcal{D}_h(\xi)$ 
by using an arbitrary signal $X^t$.
Fano's inequality is usually written
\begin{equation}
H(\theta_w\Vert X^t)
\leq 
h\left( \mathbbm{P}_w( \theta_w\not=\tilde\theta(\hat\theta_t) )\right)
+\mathbbm{P}_w( \theta_w\not=\tilde\theta(\hat\theta_t) )\log( D_h(\xi)-1).
\label{fano}
\end{equation}
where $h(x)=x\log \frac{1}{x}+(1-x)\log\frac{1}{1-x}\leq\log2$.
This says it is only possible to make really accurate predictions if $H(\theta_w\Vert X^t)$ is close to zero, so you observe signals highly correlated with what you want to predict.
As $\theta_w$ has a uniform distribution on $\mathcal{D}_h(\xi)$ under $\mathbbm{P}_w$, the left of this inequality can be rewritten
$$
H(\theta_w\Vert X^t)
=
\sum_{\theta\in \mathcal{D}_h(\xi)}b\log\frac{1}{b}-I(\theta_w, X^t)
=
\log D_h(\xi)-I(\theta_w, X^t)
$$
Here $I$ denotes the mutual information between two random variables.%
\footnote{The mutual information between two random variables is the Kullback-Leibler divergence between their joint distribution and the product of their marginals: 
$I(Y,Z):=K( P_{YZ}\Vert P_Y \otimes P_Z)$.}
Combining this with an upper bound on the RHS of (\ref{fano})
we get
$$
\log D_h(\xi)-I(\theta_w, X^t)\leq \log 2+\mathbbm{P}_w( \theta_w\not=\tilde\theta(\hat\theta_t) )\log D_h(\xi),
$$
or
$$
\mathbbm{P}_w( \theta_w\not=\tilde\theta(\hat\theta_t) )\geq 1-\frac{I(\theta_w,X^t)+\log 2}{\log D_h(\xi)}.
$$
This bound depends on the cardinality of the set of predictions $D_h(\xi)$ and on the mutual information $I(\theta_w, X^t)$ between the prior on $\mathcal{D}_h(\xi)$ and the measure, $\mathbbm{P}_w$, it induces on $X^n$.

The next step is to derive an upper bound for the mutual information term above.
(Recall that $p_w(X^t):=\sum_{\theta\in \mathcal{D}_h(\xi)} b p_\theta(X^t)$ denoted the marginal density of history $X^t$ under $\mathbbm{P}_w$.) 
\begin{align*}
I(\theta_w,X^t)
&=
\sum_{\theta\in \mathcal{D}_h(\xi)}b \int_{\mathcal{X}^t}  p_\theta(X^n)\log\frac{bp_\theta(X^t)}{b p_w(X^t)}\, \nu(dX^t)
\\
&=
\sum_{\theta\in \mathcal{D}_h(\xi)}b K ( p_\theta(X^t)\Vert p_w(X^t) )
\\
&\leq
\sum_{\theta\in \mathcal{D}_h(\xi)}b K ( p_\theta(X^t)\Vert \tilde q(X^t) )\qquad\forall \,  \tilde q(X^t)
\\
&\leq
\max_{\theta\in \mathcal{D}_h(\xi)} K ( p_\theta(X^t)\Vert \tilde q(X^t) )\qquad\forall \, \tilde q(X^t)
\end{align*}
The first inequality follows as $\tilde q=p_w=\sum_\theta b p_\theta$ minimizes the expected divergence in the third line. 
To see this notice that $\sum_\theta\omega K(p_\theta\Vert \tilde q)=B-\int p_w(X) \log \tilde q(X)\, \nu(dX)$,
where $B$ is independent of $\tilde q$, and the density $\tilde q=p_w$ maximizes the final term.
The final inequality bounds a convex sum by its largest term.

In the final step we use Lemma \ref{sb} to bound the divergence above.
Choose $\tilde q=q$ of Lemma \ref{sb} to be concentrated
on  $G_K(\varepsilon)\subset\Theta$, the set of centers of the covering set
defined just before Proposition \ref{up}.
This choice implies that $\min_{\theta_j\in G}K(p_\theta\Vert p_{\theta_j})\leq\varepsilon^2$ for all $\theta\in\Theta$.
Applying the bound of Lemma 1 to the above in this case gives 
$$
I(\theta_w,X^t)\leq \log N_K(\varepsilon) + t\varepsilon^2 
$$
Linking together all the  bounds we have found then gives
$$
\inf_{\hat\theta_t}\sup_{\theta} \mathbbm{P}_\theta\left( h(\theta,\hat\theta_t)\geq \xi/2\right)
\geq
1-\frac{I(\theta_w,X^t)+\log 2}{\log D_h(\xi)}
\geq
1-\frac{\log N_K(\varepsilon) + t\varepsilon^2 +\log 2}{\log D_h(\xi)}.
$$
Hence
$$
\inf_{\hat\theta_t}\sup_{\theta} E_\theta\left( h^2(\theta,\hat\theta_t) \right)
\geq\frac{\xi^2}{4}
\left[1-\frac{\log N_K(\varepsilon) + t\varepsilon^2 +\log 2}{\log D_h(\xi)}\right]
$$

Now we find a lower bound on $D_h(\xi)$.
First, we know that $D_h(\xi)\geq N_h(\xi)$ by the properties of covering and packing described in Section \ref{rff}.
The distances $K(p_\theta\Vert p_{\theta'})$ and $h(p_\theta,p_{\theta'})$
can be mutually bounded (see Footnote \ref{rf}) and $d_K(\theta,\theta'):=\sqrt{K(p_\theta\Vert p_{\theta'})}$, which implies that
$ h(p_\theta,p_{\theta'})\geq d_K(\theta,\theta')/\sqrt{2}M$.
Therefore, every $\xi$-ball in distance $h$ is contained in a $\sqrt{2}M\xi$-ball in the distance $K$ and so
$N_h(\xi)\geq N_K(\sqrt{2}M\xi)$.
Combining these lower bounds we have shown $D_h(\xi)\geq N_h(\xi)\geq N_K(\sqrt{2}M\xi)$ and therefore that
$$
\inf_{\hat\theta_t}\sup_{\theta} E_\theta\left( h^2(\theta,\hat\theta_t) \right)\geq\frac{\xi^2}{4}
\left[1-\frac{\log N_K(\varepsilon) + t\varepsilon^2 +\log 2}{\log N_K(\sqrt{2}M\xi)}\right].
$$

\end{proof}

\subsection{Proof of Proposition \ref{tig}}

\begin{proof}
The first step is to get good upper and lower bounds on the metric entropy terms that appear in the propositions. 
To do this we show that the $K$ distance can be well approximated by  the $L_2$ norm.%
\footnote{For $p\geq1$ the $L_p$ norm is  $\Vert x\Vert_p:=(\int |x|^p \, d\nu )^{1/p}$.}
Once this is completed we use these bounds on the entropy  to make the bounds on the learning efficiency as tight as possible.

An upper bound on $K(p\Vert q)$ can be found using the following chain of inequalities: 
$$
K(p\Vert q)\leq \log \int p^2q^{-1}d\nu \leq \int\frac{(p-q)^2}{q}d\nu\leq M\int(p-q)^2d\nu=M\Vert p-q\Vert^2_2.
$$
The first inequality follows from Jensen's inequality.
The second inequality holds as $x-1\geq\log x$. 
The final inequality applies the lower bound on the densities
in $\mathcal{P}_\Theta^{\alpha,n,\lambda}$.

To get a lower bound on $K(p\Vert q)$ we use the  chain of inequalities:
\begin{align*}
\Vert p-q\Vert_2^2
&=
\int (\sqrt{p}-\sqrt{q})^2(\sqrt{p}+\sqrt{q})^2\,d\nu
\leq 4M h^2(p,q)\leq 4MK(p\Vert q).
\end{align*}
The first inequality just applies the upper bound on $p_\theta$ and 
the final inequality follows from applying $\log x\leq 2(\sqrt{x}-1)$ to $x=q/p$.

Combining the upper and lower bounds on $K(p\Vert q)$ above, we get that
$$
\frac{1}{2\sqrt{M}}\Vert p-q\Vert_2\leq \sqrt{K(p\Vert q)}\leq \sqrt{M}\Vert p-q\Vert_2.
$$
This says that on $\mathcal{P}_\Theta^{\alpha,n,\lambda}$ the $L_2$ norm is equivalent to the $\sqrt{K}$ distance.
Every $\varepsilon$-ball in $\sqrt{K}$ contains a $\varepsilon/\sqrt{M}$-ball in $L_2$ and is itself contained in a $2\sqrt{M}\varepsilon$-ball in $L_2$. This implies that 
\begin{equation}
\log N_2(2\sqrt{M}\varepsilon)\leq \log N_K(\varepsilon)\leq \log N_2(\varepsilon/\sqrt{M}).
\label{hj}
\end{equation}
Where $\log N_2(\varepsilon)$ is the metric entropy of $\mathcal{P}_\Theta^{\alpha,n,\lambda}$ in the $L_2$ norm and $\log N_K(\varepsilon)$ is the metric entropy of $\mathcal{P}_\Theta^{\alpha,n,\lambda}$
in the $\sqrt{K}$ distance.
By \cite{Lorentz66} Theorem 10, the $L_2$ metric entropy of this space satisfies
$\log N_2(\varepsilon) = O( \varepsilon^{-1/r})$,
where $r=\alpha+n$.
Hence there exists $g<G$ and a $\bar\varepsilon$ such that for all $\varepsilon<\bar\varepsilon$
$$
g\varepsilon^{-1/r} \leq \log N_2(\varepsilon)\leq G'\varepsilon^{-1/r}.
$$
If these upper and lower bounds are applied to (\ref{hj}) we have that 
\begin{equation}
g'\varepsilon^{-1/r} \leq \log N_K(\varepsilon)\leq G'\varepsilon^{-1/r},
\label{we}
\end{equation}
where $g'=g(2\sqrt{M})^{-1/r}$ and $G'=GM^{1/2r}$. 
This completes the first step in the argument.

We now use the upper bound on $\log N_K$ in (\ref{we}) and some elementary calculus to find the tightest upper bound in (\ref{yb1}).
Clearly
\begin{equation}
\inf_\varepsilon \varepsilon^2+t^{-1}\log N_K(\varepsilon)
\leq
\min_\varepsilon \varepsilon^2+t^{-1}\frac{G'}{\varepsilon^{1/r}} 
=
(2r+1)\left(\frac{G'}{2r t}\right)^{\frac{2r}{1+2r}}
\label{za}
\end{equation}
By Proposition \ref{up} this provides an upper bound on learning efficiency
$$
\inf_{\hat\theta_t\in S_t}\sup_{\theta\in\Theta}
E_\theta\left( h^2(p_\theta , p_{\hat\theta_t})\right)
\leq C t^{\frac{-2r}{1+2r}},
$$
for some constant $C$.

If the upper bound of (\ref{za}) is substituted into (\ref{dd}) we get
$$
\inf_{\hat\theta_t\in S_t}\sup_{\theta\in\Theta}
E_\theta\left( h^2(p_\theta , p_{\hat\theta_t})\right)
\geq 
\frac{\xi^2}{4}
\left[1-\frac{t(2r+1)\left(\frac{G'}{2r t}\right)^{\frac{2r}{1+2r}} +\log 2}{\log N_K(\sqrt{2}M\xi)}\right].
$$
To bound this from below we need to apply the lower bound (\ref{hj}) on $\log N_K$ in the denominator above.
Hence
$$
\sup_{\xi>0}
\frac{\xi^2}{4}
\left[1-\frac{t(2r+1)\left(\frac{G'}{2r t}\right)^{\frac{2r}{1+2r}} +\log 2}{\log N_K(\sqrt{2}M\xi)}\right]
\geq
\max_{\xi>0} 
\frac{\xi^2}{4}
\left[1-\frac{t(2r+1)\left(\frac{G'}{2r t}\right)^{\frac{2r}{1+2r}} +\log 2}{g''\xi^{-1/r} }\right]
$$
where $g''=g'(\sqrt{2}M)^{-1/r}$.
Now choose $\xi$ so that $g''\xi^{-1/r}=2t(2r+1)\left(\frac{G'}{2r t}\right)^{\frac{2r}{1+2r}}$.
For this choice of $\xi$ we get
\begin{align*}
\max_{\xi>0} 
\frac{\xi^2}{4}
\left[1-\frac{t(2r+1)\left(\frac{G'}{2r t}\right)^{\frac{2r}{1+2r}} +\log 2}{h''\xi^{-1/r} }\right]
&\geq
\frac{\xi^2}{8}-\frac{\log 2}{g''}\xi^{2+r^{-1}}
\\
&=t^{\frac{-2r}{1+2r}}
\left(
\frac{(g'')^{r+1/2}G'}{ (2r+1)^{r+1/2}2r }
\right)^{\frac{4r}{2r+1}}
-
O(t^{-1})
\end{align*}
By Proposition \ref{down}, therefore we have a lower bound on learning efficiency.
$$
\inf_{\hat\theta_t\in S_t}\sup_{\theta\in\Theta}
E_\theta\left( h^2(p_\theta , p_{\hat\theta_t})\right)
\geq 
t^{\frac{-2r}{1+2r}}
C'
-
O(t^{-1})
$$

\end{proof}

\subsection{Proof of Proposition \ref{lo}}

\begin{proof}
 We start by showing that the manager's objective can be bounded below by the measure of learning efficiency we use.
Consider the following chain of inequalities
\begin{align*}
\int_{[0,1]} (\sqrt{e}-\sqrt{f_\theta})^2\,d\nu
&
=
2-2\int_{[0,1]} \sqrt{e f_\theta}\,d\nu
\\
&
\geq
2-2\left(\frac{1}{2} \int_{[0,1]} e^\alpha f_\theta^{1-\alpha}\,d\nu+\frac{1}{2} \int_{[0,1]} e^{1-\alpha} f_\theta^{\alpha}\,d\nu\right)
\\
&
\geq
1- \int_{[0,1]} e^\alpha f_\theta^{1-\alpha}\,d\nu
\end{align*}
The first inequality holds as $\int e^\alpha f^{1-\alpha}\,d\nu$ is a convex function of $\alpha\in[0,1]$ and the final inequality follows as this function is bounded above by one. 
Comparing the ends of the above chain of inequalities we get
$1-h^2(e,f_\theta)\leq \int_{[0,1]}e^\alpha f_\theta^{1-\alpha}d\nu$ for all $ \alpha\in[0,1]$.
By taking a tangent to the convex function at $x=1$ we have $E(x^{1/(1-\beta)})\geq (E(x)-\beta)/(1-\beta)$ and so
\begin{align*}
E_\theta\left( \left(\int_{[0,1]} e^{1-\beta}f_\theta^\beta\,d\nu\right)^{\frac{1}{1-\beta}}\right)
&
\geq
E_\theta\left( \left( 1-h^2(e,f_\theta)\right)^{\frac{1}{1-\beta}}\right)
\\
&
\geq
1-\frac{1}{1-\beta}E_\theta (h^2(e,f_\theta) )
\end{align*}

Once we have this lower bound it is a simple matter to apply  the upper bound of Proposition \ref{tig} to the learning efficiency to get a lower bound on the manager's payoff.
We have that
\begin{align*}
1\geq
&\sup_{e_t}\inf_{f_\theta\in \mathcal{P}_\Theta^{\alpha,n,\lambda}}
E_\theta\left( \left(\int_{[0,1]} e_t^{1-\beta}f_\theta^\beta\,d\nu\right)^{\frac{1}{1-\beta}}\right)
\\
&\geq
1-\frac{1}{1-\beta}
\inf_{e_t}\sup_{f_\theta\in \mathcal{P}_\Theta^{\alpha,n,\lambda}}E_\theta (h^2(e_t,f_\theta) )
\\
&\geq
1-C t^{\frac{-2r}{1+2r}}.
\end{align*}
To get an upper bound we rewrite the equality $1-h^2(e,f)=\int\sqrt{ef}\,d\nu$ as
$$
-2\log(1-\frac{1}{2}h^2(e,f))=\frac{1}{\beta-1}\log\int e^\beta f^{1-\beta}d\nu
$$
for $\beta=1/2$. By, for example \cite{van14}, the right is increasing in $\beta$ on $[0,1]$ and so for all $\beta\in[1/2,1)$ we have
\begin{align*}
-2\log(1-\frac{1}{2}h^2(e,f))
&\leq 
\frac{1}{\beta-1}\log\int e^{1-\beta} f^\beta d\nu
\\
2\log(1-\frac{1}{2}h^2(e,f))
&\geq 
\log\left(\int e^{1-\beta} f^{\beta}d\nu\right)^{\frac{1}{1-\beta}}
\\
(1-\frac{1}{2}h^2(e,f))^2
&\geq 
\left(\int e^{1-\beta} f^{\beta}d\nu\right)^{\frac{1}{1-\beta}}
\\
1-\frac{1}{2}h^2(e,f)
&\geq 
\left(\int e^{1-\beta} f^{\beta}d\nu\right)^{\frac{1}{1-\beta}}.
\end{align*}
This inequality holds for all $\beta\in(1/2,1)$.
If we swap $e$ an $f$ in the above we get
$$
(1-\frac{1}{2}h^2(e,f))^2
\geq 
\left(\int e^{\beta'} f^{1-\beta'}d\nu\right)^{\frac{1}{1-\beta'}},
\qquad \forall \beta'\in(0.5,1).
$$
Then letting $\beta=1-\beta'$ we get
$$
(1-\frac{1}{2}h^2(e,f))^2
\geq 
\left(\int e^{1-\beta} f^{\beta}d\nu\right)^{\frac{1}{\beta}},
\qquad \forall \beta\in(0,0.5).
$$
Or
$$
(1-\frac{1}{2}h^2(e,f))^{\frac{2\beta}{1-\beta}}
\geq 
\left(\int e^{1-\beta} f^{\beta}d\nu\right)^{\frac{1}{1-\beta}},
\qquad \forall \beta\in(0,0.5).
$$
Assuming $\beta\geq1/3$ the left hand side above is smaller than $1-\frac{1}{2}h^2$.
Thus we have extended the previous inequality to this case.
Taking expectations we then get
$$
1-\frac{1}{2}E_\theta \left( h^2(e_t,f_\theta)\right)
\geq 
E_\theta\left( \left(\int_{[0,1]} e_t^{1-\beta} f_\theta^{\beta}d\nu\right)^{\frac{1}{1-\beta}}\right),
\qquad \forall \beta\in[1/3,1).
$$
By applying the bound in Proposition \ref{tig} we get for any $\beta\geq1/3$
\begin{align*}
1-C't^{\frac{-2r}{1+2r}}-O(t^{-1})
&\geq
1-\inf_{e_t}\sup_{f_\theta} \frac{1}{2}E_\theta \left( h^2(e_t,f_\theta)\right)
\\
&\geq 
\sup_{e_t}\inf_{f_\theta} E_\theta\left( \left(\int_{[0,1]} e_t^{1-\beta} f_\theta^{\beta}d\nu\right)^{\frac{1}{1-\beta}}\right).
\end{align*}

\end{proof}

\subsection{Proof of Proposition \ref{fgg}}

\begin{proof}
The first step in the proof is derive a relationship between the maximum and the integral of a non-negative H\"older function.
Suppose the functions $f:[0,1]\rightarrow\mathbbm{R}$ and $g:[0,1]\rightarrow\mathbbm{R}$  satisfy the conditions
\begin{equation}
|f(x)-f(x+h)|\leq\lambda|h|^\alpha,
\qquad
|g(x)-g(x+h)|\leq\lambda|h|^\alpha;
\label{hh}
\end{equation}
for all $x,x+h\in[0,1]$.
Then by the triangle inequality, the function $\delta(x):=|f(x)-g(x)|$ satisfies $|\delta(x)-\delta(x+h)|\leq2\lambda|h|^\alpha$.
If $\delta(0)=0$ and $\delta(.)$ is increasing then $\delta(x)\leq 2\lambda x^\alpha$ and for all $y\leq x$ we have:
$$
\delta(x)-\delta(y)\leq 2\lambda (x-y)^\alpha
$$
This gives the lower bound $\delta(y)\geq \max\{\delta(x)- 2\lambda (x-y)^\alpha,0\}$.
Hence
$$
\int_0^x \delta(y) \,dy 
\geq 
\int_{x-(\frac{\delta(x)}{2\lambda})^{1/\alpha}}^x \delta(x)- 2\lambda (x-y)^\alpha \,dy
=
\delta(x)^{\frac{1+\alpha}{\alpha}}\frac{\alpha(2\lambda)^{-\frac{1}{\alpha}}}{\alpha+1}.
$$

If $f,g\in \mathcal{P}_\Theta^{\alpha,n,\lambda}$ they both integrate to one and are continuous so there must be an $x\in [0,1]$ where $|f-g|=0$.
Hence, the inequality above implies that if $f$ and $g$ satisfy the above H\"older condition then
$$
\Vert f-g\Vert_1
\geq 
\frac{\alpha(2\lambda)^{-\frac{1}{\alpha}}}{\alpha+1} \Vert f-g\Vert_\infty^{\frac{1+\alpha}{\alpha}}.
$$
Notice that even when $n>1$ the functions $f,g\in \mathcal{P}_\Theta^{\alpha,n,\lambda}$  satisfy the condition (\ref{hh}) with $\alpha=1$, because their derivatives are bounded by $\lambda$.
(Or one could apply the Hardy-Littlewood-Landau inequality to get the same conclusion.)
Thus when $n>1$ we take $\alpha =1$ in the above inequality.

The second step in the proof is to use the inequality $2h(p,q)\geq \Vert p-q\Vert_1$  \cite[Appendix B]{GhosalVDV17} to get that
$$
h^2(p,q)
\geq 
\frac{\alpha^2(2\lambda)^{-\frac{2}{\alpha}}}{4(\alpha+1)^2} 
\Vert p-q\Vert_\infty^{\frac{2(1+\alpha)}{\alpha}},
\qquad
(n>0\Rightarrow\alpha=1).
$$
From our upper bound in Proposition \ref{tig} we have that 
\begin{align*}
C t^{\frac{-2\alpha}{1+2\alpha}}
&\geq
\frac{\alpha^2(2\lambda)^{-\frac{2}{\alpha}}}{4(\alpha+1)^2}
\inf_{\hat\theta_t\in S_t}\sup_{\theta\in\Theta}
E_\theta\left( \Vert p_\theta-p_{\hat\theta_t}\Vert_\infty^{\frac{2(1+\alpha)}{\alpha}} \right),
\qquad n=0;
\\
C t^{\frac{-2r}{1+2r}}
&\geq
\frac{\lambda^{-2}}{64}
\inf_{\hat\theta_t\in S_t}\sup_{\theta\in\Theta}
E_\theta\left( \Vert p_\theta-p_{\hat\theta_t}\Vert_\infty^{4} \right),
\qquad \qquad \qquad n>0.
\end{align*}
From Jensen's inequality  and combining terms into a new constant we get
\begin{align}
C t^{\frac{-\alpha^2}{(1+2\alpha)(1+\alpha)}}
&\geq
\inf_{\hat\theta_t\in S_t}\sup_{\theta\in\Theta}
E_\theta\left( \, \Vert p_\theta-p_{\hat\theta_t}\Vert_\infty \, \right),
\qquad n=0,
\label{mk}
\\
C t^{\frac{-r}{2(1+2r)}}
&\geq
\inf_{\hat\theta_t\in S_t}\sup_{\theta\in\Theta}
E_\theta\left( \, \Vert p_\theta-p_{\hat\theta_t}\Vert_\infty \, \right),
\qquad n>0.
\end{align}

Recall that we defined, as a benchmark, the best achievable payoff for the DM if they knew the state $\theta$,
$$
M_\theta:=\max_{D\in\mathcal{D}}d^{-1} \int \mathbbm{1}_{D}p_\theta\, d\nu.
$$
Let $\hat D_t$ be the solution to the problem of finding the alternatives that
maximize the estimate function
$$
d^{-1}\int \mathbbm{1}_{\hat D_t}p_{\hat\theta_t} \, d\nu
=
\max_{D\in\mathcal{D}} d^{-1}\int \mathbbm{1}_{D}p_{\hat\theta_t} \, d\nu.
$$
By definition of the $\Vert.\Vert_\infty$ norm, for all $x$  
$$
p_{\hat\theta_t}(x)+\Vert p_\theta-p_{\hat\theta_t}\Vert_\infty\geq p_\theta(x)
\geq
p_{\hat\theta_t}(x)-\Vert p_\theta-p_{\hat\theta_t}\Vert_\infty.
$$
But this implies bounds on the true optimal payoff
$$
d^{-1}\int \mathbbm{1}_{\hat D_t}p_{\hat\theta_t} \, d\nu
+
d^{-1}\Vert p_\theta-p_{\hat\theta_t}\Vert_\infty
\geq 
M_\theta
\geq
d^{-1}\int \mathbbm{1}_{\hat D_t}p_{\hat\theta_t} \, d\nu
-
d^{-1}\Vert p_\theta-p_{\hat\theta_t}\Vert_\infty.
$$
Also as $\hat D_t$ is suboptimal for the true state we have that
$$
M_\theta
\geq
d^{-1}\int \mathbbm{1}_{\hat D_t}p_{\theta} \, d\nu
\geq
d^{-1}\int \mathbbm{1}_{\hat D_t}p_{\hat\theta_t} \, d\nu
-
d^{-1}\Vert p_\theta-p_{\hat\theta_t}\Vert_\infty.
$$
Combining these families of inequalities we get
$$
0\leq M_\theta -d^{-1}\int \mathbbm{1}_{\hat D_t}p_{\theta} \, d\nu
\leq
d^{-1}\Vert p_\theta-p_{\hat\theta_t}\Vert_\infty.
$$
Then an application of (\ref{mk}) to the right of this gives
$$
0
\leq
\inf_{\hat D_t}\sup_{\theta\in\Theta}
E_\theta\left( \, 
M_\theta -d^{-1}\int \mathbbm{1}_{\hat D_t}p_{\theta} \, d\nu
 \, \right)
\leq
d^{-1}C t^{\frac{-\alpha^2}{(1+2\alpha)(1+\alpha)}}.
$$
Hence there is a strategy $\hat D_t$ for the DM so that
for all $p_\theta$
$$
E_\theta\left( \, 
M_\theta -d^{-1}\int \mathbbm{1}_{\hat D_t}p_{\theta} \, d\nu
 \, \right)
 \leq
 d^{-1}C t^{\frac{-\alpha^2}{(1+2\alpha)(1+\alpha)}}
$$
\end{proof}

\end{document}